\documentstyle[12pt,cmb]{article}

\newcommand{\dr}{\delta_R}

\newcommand{\so}{{\cal S}}

\newcommand{\beq}{\begin{eqnarray}}
\newcommand{\eeq}{\end{eqnarray}}

\begin{document}
\begin{flushright}
CfPA/96-th-12
\end{flushright}
\heading{MICROWAVE ANISOTROPIES FROM RANDOM SOURCES}

\author{Pedro G. Ferreira,}{Center for Particle Astrophysics\\
University of California\\ Berkeley,  CA 94720-7304\\ USA.}{\quad}

\begin{abstract}{\baselineskip 0.4cm
I report on recent developments in the theory of
cosmic background radiation perturbations.
I describe ways of modeling alternatives to the canonical  
Gaussian theories within the
standard framework of cosmological perturbation theory. Some 
comments are made on using these techniques to resolve
 the uncertainties in theories of structure formation with
topological defects. (To appear in the proceedings of the XXXIth
Moriond meeting, ``Microwave Background Anisotropies'')
\\}
\end{abstract}

The past few years have seen a remarkable development in both the 
experimental and theoretical understanding of the cosmic backround
radiation (CBR). Following the tremendous success of COBE in mapping out
the sky to $10^{\rm o}$ resolution and showing us that, indeed these
fluctuations do exist, a flurry of results have come out of
groundbased and balloon borne, degree scale experiments. Although it
seems premature to make strong claims about the fine features
of the current results, it is clear that we are very rapidly
progressing to a very detailed knowledge of what the CBR looks like
on all scales.
Driven by the wealth of experimental data which will be available, 
theorists have been developing numerical and analytic tools which
help us understand what impact these measurements have on our world 
model. 

Almost all efforts have focused on a restricted set of models
where perturbations are set up in the initial conditions and are
elements of a Gaussian ensemble. 
A number of groups have developed numerical algorithms
for evolving perturbations in radiation in an expanding universe
(see Seljak's contribution in this proceeding
 for an exciting recent development) and it is now possible
to calculate the angular power spectrum of the CBR for these models
with a precision
of smaller than a percent on all scales.

The determination of the statistical properties of the CMB
anisotropies involves evolving the linearized Boltzman equation \cite{Boltz}.
One can identify various regimes and for the purpose of this 
proceeding I shall focus on the pre-recombination
era where radiation behaves as a tightly coupled fluid with an
effective equation of state $P={1 \over 3}\rho$. If we make the
further simplification of assuming that recombination happens in the 
radiation era, then it suffices to consider the following set of
equations \cite{Def}:
\beq
{\ddot h}+{1 \over \eta}{\dot h}+{6 \over \eta^2}\delta_R&=&-8\pi G
\so_+\nonumber \\
{\ddot \delta}_R+{k^2\over 3}\delta_R+{2 \over 3}{\ddot h}&=&0 \nonumber
\\
{\ddot h}^S+{2\over \eta}{\dot h}^S+4\tau_{00}&=&8\so^S \label{minib}
\eeq
where $\delta_R$ is the radiation density contrast, we have decomposed
the scalar spatial tensors into $A_{ij}={1 \over 3}A\delta_{ij}+({\hat
k}_i{\hat k}_j-\delta_{ij})A^S$, $\eta$ is conformal time and 
$\tau_{00}=\so_{00}+{3 \over {4\eta^2}}\dr-{1 \over {4\eta}}{\dot h}$
is the ``pseudo-energy'' of the system. $\so_{\mu\nu}$ ($\so_+=
\so_{00}+\so_{ii}$) is the
energy-momentum tensor of an external source with its own set of
statistical properties; we shall assume that its dominant form of
self interaction is non-gravitational. Topological defects are
examples of such sources. If we are interested in the angular
power spectrum, one must project this set of perturbations forward; for
example if one coniders the monopole at last scattering then one
obtains
\beq
C_l={1 \over {16}}\int_0^\infty k^2dk<|\delta_R-{{2{\ddot h}^S}\over k^2}|^2>
j^2_l(k(\eta_0-\eta_*)) \nonumber
\eeq

The canonical picture of structure formation relies on a few technical
assumptions which simplify calculations tremendously. Firstly
perturbations are set up in the early universe, deep in the radiation
era (at $\eta_i$). For example {\it Inflation} will imprint a set of adiabatic
perturbations on superhorizon scales at the Planck time. The
subsequent evolution of these perturbations is studied with equations
(\ref{minib}) where we set all $\so$'s to $0$. Note that this means
that all variables, as a function of time, are homogeneous functions of
degree one in the initial conditions; heuristically this means that,
for example
\beq
\dr(k,\eta)\simeq T_R(k;\eta,\eta_i)\dr(k,\eta_i) \nonumber
\eeq
Secondly the initial set of perturbations is a realization
from a Gaussian ensemble with 0 mean, i.e. to describe
the statistics of the ensemble {\it completely} 
at $\eta_i$ it suffices to specify $<|\dr(k,\eta_i)|^2>$.  These two 
technical assumptions lead to a remarkably simple algorithm for studying
the statistics of perturbations today. One picks as initial conditions
the square root of the initial variances of the variables, evolves them
forward deterministically using equations (\ref{minib}) and squares the
result. 

We could however consider a much more general set of perturbations,
with non-zero $\so$ and initial conditions with more complex
statistics. A restricted set of examples have been studied in this
class: if one assumes that the universe underwent a symmetry breaking
phase transition, and that the symmetry breaking pattern satisfies
a certain set of conditions, than it is possible that topological
relics could have formed. For structure formation, the interesting
examples are {\it cosmic strings} and textures. With the aim of going
through the same sort of analysis as we have just performed for Gaussian
theories let us focus on cosmic
strings.

Two important features arise: Firstly, if we assume that universe
was homogeneous and isotropic and make the natural assumption that
all processes during the phase tranistion were causal, than the
initial conditions to equation (\ref{minib}) satisfy a 'causality
constraint'. In particular $\tau_{00}\propto k^2$ on superhorizon
scales. Note that this implies a delicate cancellation between the
``source'' (which, as we shall see, is $\propto k^0$), the other
fluids and the gravitational field. Secondly, we must know the statistics of the source. Unlike
the case of primordial, Gaussian fluctuations, here we have to
describe the time evolving statistics. We must characterize these
sources as an ensemble of histories, which satisfy a certain set of
properties. In the case of cosmic strings we have\\
i)\ \ \ \ \ \ \  $<\so>\simeq 0$\\
iii) causality enforces the sources to have no correlations on
superhorizon scales; this means 
\beq
<\so_{00}({\bf x},\eta)\so_{00}({\bf
x},\eta)>=0 \ \ \ \mbox{ if \ \ $|{\bf x}-{\bf x}'|>\eta+\eta'$}\nonumber
\eeq
 which implies
\beq
<|\so_{00}(k,\eta)|^2>\propto k^0 \ \ \ \mbox{for \ \  $k\eta \ll 1$.}
\nonumber
\eeq
iii) $\so_{00}$ satisfies ``active'' scaling so that
\beq
<|\so_{00}(k,\eta)|^2>\simeq{1 \over {\eta^{1 \over 2}}}f(k\eta)\nonumber
\eeq
This means that perturbations will be seeded with a constant amplitude
at order of the horizon size, inducing ``quasi'' scale invariance
``actively''. This condition may not be strictly enforced in certain
eras; cosmic strings deviate mildly from it in the radiation/matter
transition

\vbox to 0.4\vsize{
\includegraphics{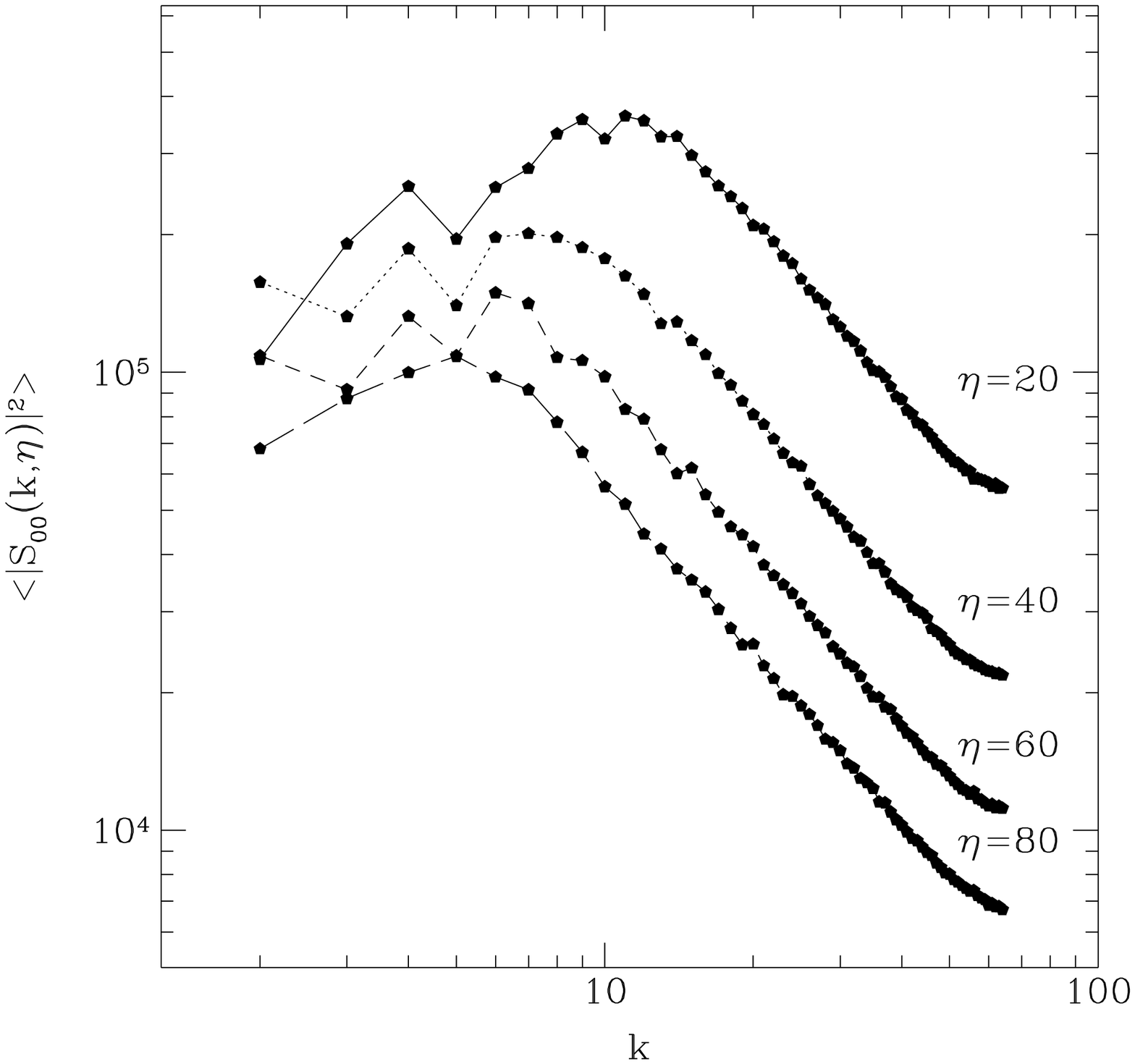}
\vfil
{
\noindent Figure 1: Equal time power spectrum for the source energy density
at different time steps, from a 3D simulation of a cosmic string 
network [4]. Note the evidence for scaling and the $k^{-1}$ behaviour,
typical of line like configurations, on small scales.
}}
\vskip .1in

iv)In the case of cosmic strings, the sources are a highly non-linear
system, i.e. there is a strong coupling between the evolution of the
different modes. As a consequence there is an active randomization
of each fourier modes in such a way that, for example
\beq
<\so_{00}(k,\eta)\so_{00}(k,\eta+\Delta)>\simeq 0 \ \ \ \ \mbox{if $\Delta>\eta_c$}\nonumber
\label{unet}
\eeq
for some coherence time $\eta_c$ (and the same can be said of
other components of $\so$). These features
of the unequal time correlation function may have striking
observational consequences. As pointed out in \cite{ACFM} the 
structure of the secondary peaks will be completely smoothed
out if the $\eta_c$ is small enough and  $\so_+$ 
has enough ``power'' on small scales.\\ 
\vbox to .1\vsize{
\includegraphics{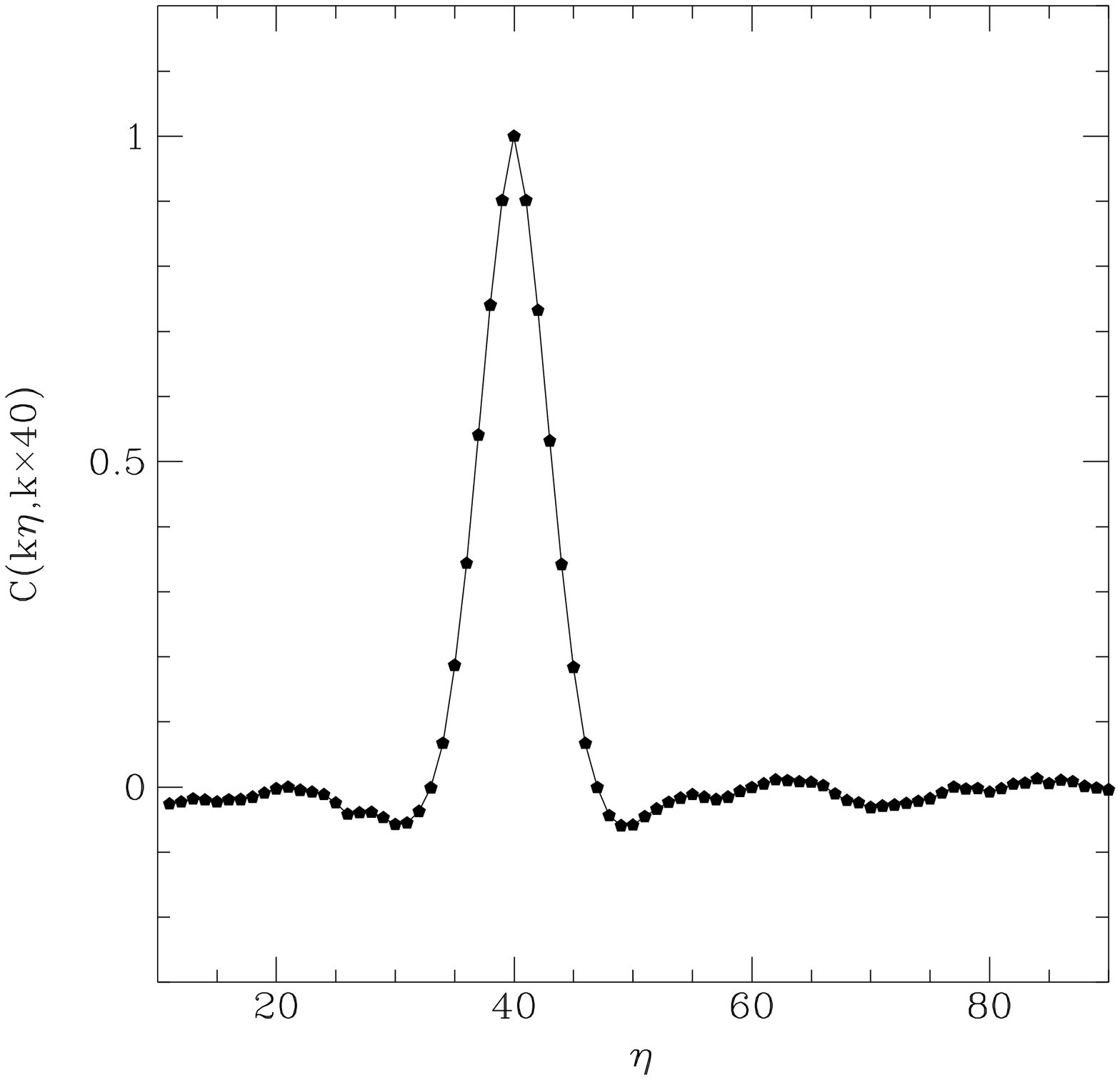}
\includegraphics{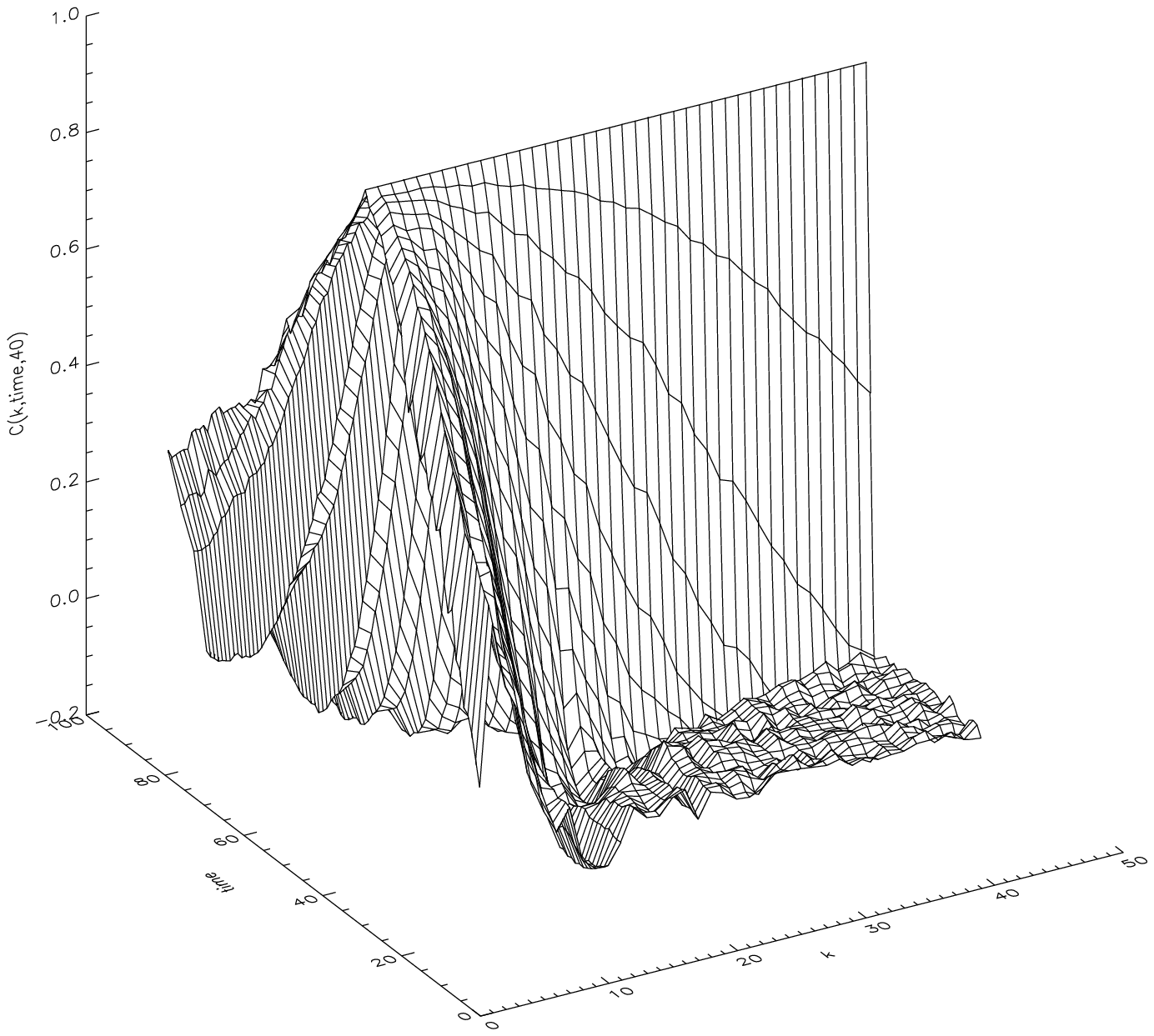}
}
\vfil
{
\noindent Figure 2: 
Plots of unequal time correlation function of $\so_{00}$ normalized to
unity at equal time, C (see \cite{PRD1,PRD2}). The right  hand side is a surface plot with one of
the entries fixed at $\eta=40$ in box units (note that $\eta_c$ is scale
dependent).
 The left hand side is a slice through
the surface plot (for one mode, $k=20$)}
\vskip .1in

I should emphazise that this class of perturbations is much more 
general than the canonical set. To pick a theory one must define
the statistics of the initial condition and of the source ensemble.
Primoridal Gaussian theories are a sub-class with Gaussian initial
conditions and $\so_{\mu\nu}=0$ which are particularly easy to study.

Having set up the system in this way we can now look at different
strategies for studying specific examples in this general class of
theories. Again I will focus on cosmic strings. Ideally one should
be able to generate an ensemble of evolving networks through
some numerical algorithm. These would supply us with $\so_{\mu\nu}$
which would go into eq (\ref{minib}) and generate a set of $\delta_R$
and $h^S$ which we could use to reconstruct 2d (or even 3d) realizations
of the CMB anisotropy. This technique has the tremendous advantage
of supplying us with {\it all} the information in CMB anisotropies
generated by strings, i.e. with a reasonable amount of realizations we
can reconstruct the full statistical distribution of anistropies.
There are serious practical problems however. Existing high-resolution
codes of cosmic string networks have a very limited dynamical range
and it is unclear whether one is actually probing the true physical
scaling regime of the network. Modified flat-space codes have the
advantage of being very fast and have a large dynamic range but it is
even less clear whether these numerical systems represent the real
world with sufficient accuracy. Also care must be had when evolving
the combined system of defects and gravity to ensure that
energy-momentum is properly consereved (for a description of the
problem see \cite{Def}).

\vbox to 0.38\vsize{
\includegraphics{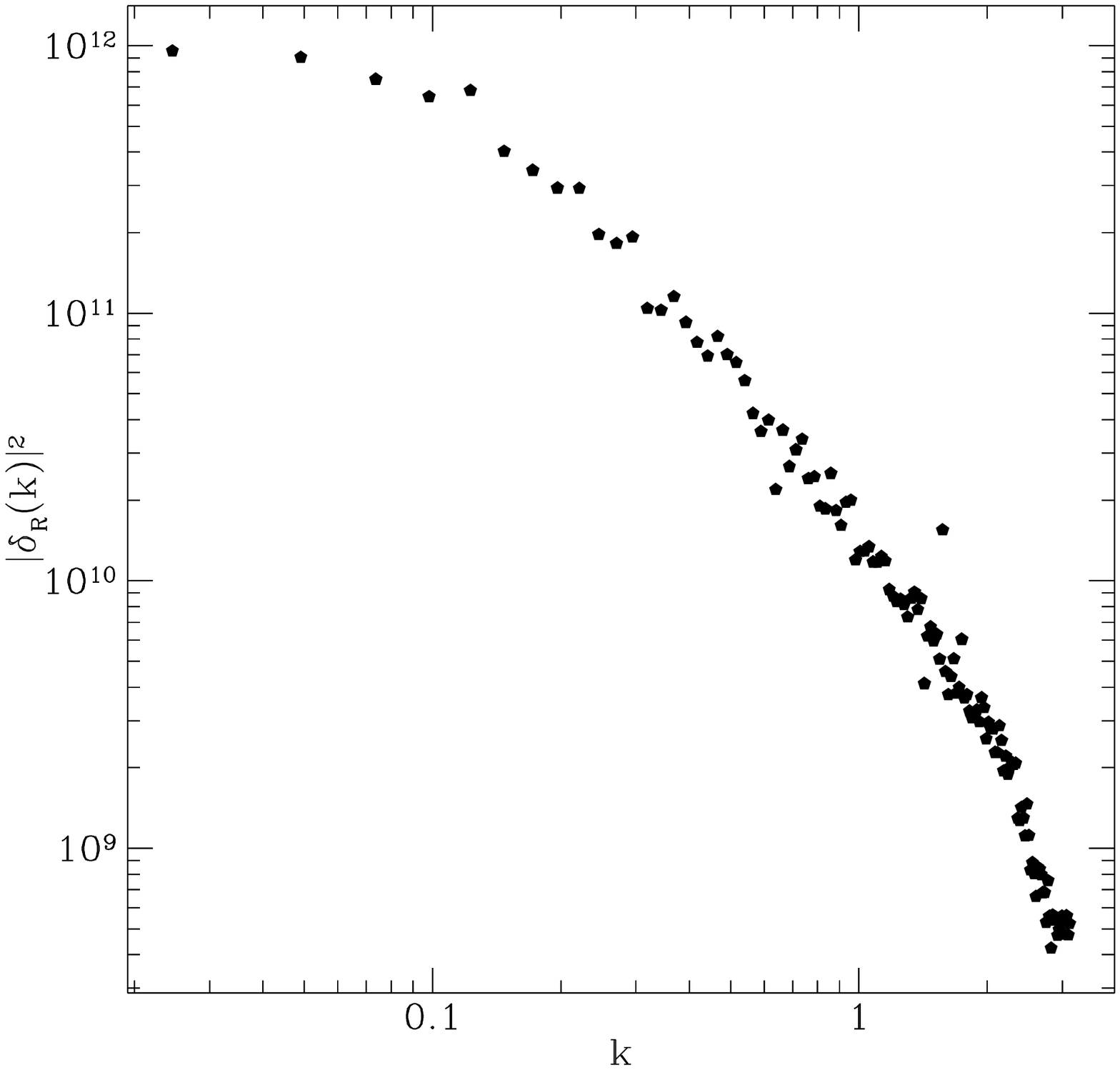}
\includegraphics{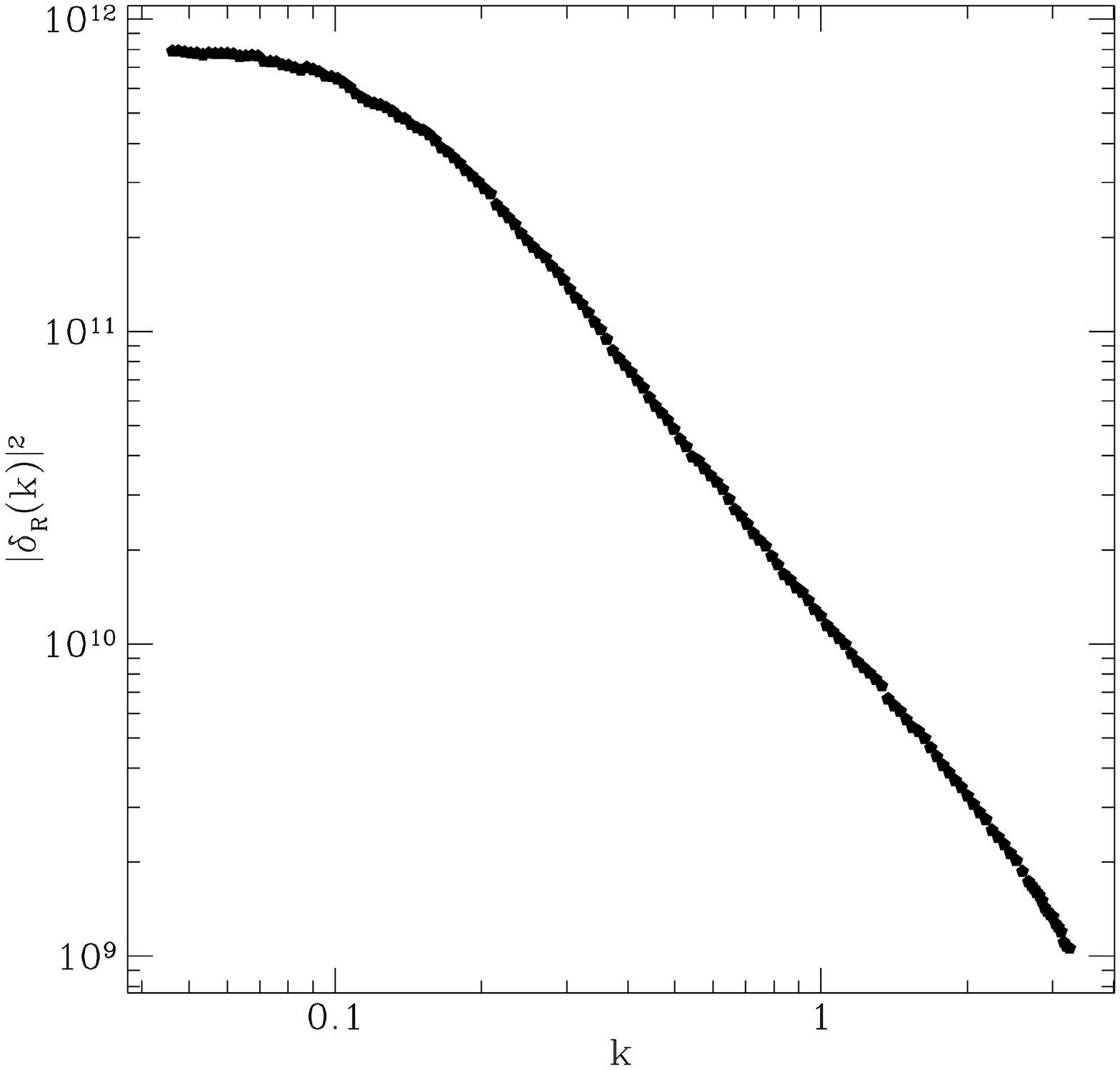}
}
\vfil
{
\noindent Figure 3: 
The power spectrum of the $\delta_R$ at last scattering using two
different techniques: on the left hand side we use a full 3d network
code to generate the sources \cite{Turok}; on the right hand side we use a random
number generator to generate a source value at each time step of
typical size $\eta_c$.}
\vskip .1in

A simplification recently suggested by Turok \cite{Turok} 
is to evolve the a defect
network and store a small set of one-dimensional fourier transforms
(along the main axis, for example), i.e. at each $\eta$ all we need
is (for example)
\beq
{\bar S}_+(x,\eta)\equiv\int dydz \so_+(x,y,z,\eta) \nonumber
\eeq
 In this way one reduces
considerably the time needed for generating each history and,
although one cannot reconstruct 2d (3d) realizations of the anisotropy
field, one retains all the statistical information of each mode. Again
one is tied down to the accuracy of the code modeling the defect
network and it is not possible to vary the phenomenological parameters
of the theory and estimate uncertainties in the observables.

One can forgo the use of a defect network if one restricts oneself to 
estimating the ensemble averaged power-spectrum. It is quadratic
in the perturbation variables and therefore quadratic in the sources.
If one generates an ensemble of $\so$ histories which have the same
two-point
correlation properties of the real network then one will obtain the
correct averaged power spectrum. Note that this is all we need to
estimate the
mean of $C_l$ but is not enough enough for the variance of $C_l$
(the ``cosmic'' variance) which may be significantly different
from the Gaussian case. A possible algorithm is the following:\\
a) evaluate the the statistical properties form a string network
simulation\\ (e.g. $<\so_+(k,\eta)\so_+(k,\eta')>$,
$<\so^S(k,\eta)\so^S(k,\eta')>$,  $<\so_+(k,\eta)\so^S(k,\eta')>$,
etc.)\\
b) divide the whole time interval over which you are going to evolve
Eq. (\ref{minib}) into subintervals of size $\eta_c$. An obvious
refinement
is to choose each time step from a distribution with mean value $\eta_c$.\\
c) use a gaussian number generator to generate
realizations of $\so_+$, $\so_{00}$ and $\so^S$ in each subinterval
with the variances from a)\\
d) feed these histories into Eq. (\ref{minib}).

The tremendous advantage is that one can sidestep the uncertainties
 in modeling the defect network. By making a realistic guess of the
uncertainties in the properties evaluated in a) one can study their
impact on the estimate of the angular power spectrum. An example
of this has been attempted (although in a different formalism) for
cosmic strings in \cite{MACF,PRD1}; it was found that current uncertainties
lead us to be uncertain about the height of the ``Doppler peak''
to within a factor of 10, that its position can be estimated to lie
between $l$ of 400 and 600 but that one can safely say that there will
be
little or NO secondary oscillations. One hopes that in this way one can
circumvent the uncertainties which have plagued the study the 
cosmological implications of topological defects during the last 20
years. One important problem that can be addressed has been reiterated
by White and Scott \cite{WS}; if one accepts the current COBE normalization
of these models to be correct then there is serious lack of power
on medium to small scales in the matter power spectrum (on
sufficiently
large scales to be difficult to justify in terms of ``bias'').
With this method it is possible to assess how serious a problem 
this is, given the current uncertainites in modeling the defect
network.

I have focused on cosmic strings where we have some intuition
(and experience) of the behaviour of the sources. However it is
possible to consider these sources as a ``phenonmenological''
ingredient in a theory of structure formation, in the same way
that Gaussian initial conditions were postulated before the
advent of inflation.

\acknowledgements{This work was done in collaboration with 
A. Albrecht, D. Coulson and J. Magueijo. I acknowledge
useful conversations with R. Durrer, W. Hu, U. Seljak, J. Silk, N. Turok and
M. White. I was supported by the Center for Particle Astrophysics,
a NSF Science and Technology Center at UC Berkeley, under Cooperative
Agreement No. AST 9120005.}

\vfill

\begin{thebibliography}{99}{\baselineskip 0.4cm
\bibitem{Boltz} Wilson M.L. and Silk J. {\it Astrophys. Jour.} {\bf
243} 14 (1981); Bond J.R. and Efstathiou G. {\it Astrophys. Jour}
{\bf 285} L45 (1984); Hu W. and Sugiyama N. {\it Astrophys. Jour}
{\bf 444} 489 (1985)
\bibitem{Def} Veeraraghavan S. and Stebbins A., {\it Astrophys. Jour}
{\bf 365} 37 (1990); Pen U.L., Spergel D.N. and Turok N. {\it Phys.
Rev. D} 692 (1994)}
\bibitem{ACFM} Albrecht A., Coulson D., Ferreira P. and Magueijo J.,
{\it Phys. Rev. Lett.} {\bf 76} 1413 (1996)
\bibitem{MACF} Magueijo J., Albrecht A., Coulson D. and Ferreira P.,
{\it Phys. Rev. Lett.} {\bf 76} 2617 (1996)
\bibitem{PRD1} Magueijo J., Albrecht A., Coulson D. and Ferreira P.,
\bibitem{PRD2} Ferreira P., Albrecht A. and Magueijo J., in preparation
 astro-ph/9605047, sub. to {\it Phys. Rev. D}
\bibitem{Turok} Turok N., {\it private communication}
\bibitem{WS} White M. and Scott D. astro-ph/9601170, sub. to {\it
Comm. in Astro.}
\end{thebibliography}
\end{document}